\documentclass[aps,prd,amsfonts,twocolumn]{revtex4}

\usepackage{color}
\usepackage{graphics,epsfig}

\usepackage{mathpazo}   

\usepackage{epstopdf}

\newcommand {\be} {\begin{equation}}
\newcommand {\ba} {\begin{eqnarray}}
\newcommand {\ee} {\end{equation}}
\newcommand {\ea} {\end{eqnarray}}

\renewcommand{\Re}{{\rm Re\,}}

\begin{document}

\preprint{WM-07-105}

\title{Rosenbluth Nonlinearity from Two-Photon Exchange}

\author{Zainul Abidin and Carl E.\ Carlson}
\affiliation{
Department of Physics, College of William and Mary, Williamsburg, VA 23187, USA}

\date{\today}

\begin{abstract}
We calculate, using a generalized parton distribution based formalism, the nonlinearity of the Rosenbluth plots that arise from hard two-photon exchange corrections that are not included in the classic calculations of the radiative corrections to electron-proton elastic scattering.
\end{abstract}

\maketitle

%
\section{Introduction}
%

Previous investigations~\cite{Guichon:2003qm,Blunden:2003sp,Chen:2004tw,Afanasev:2005mp,Blunden:2005ew,Kondratyuk:2005kk} have shown that a possible explanation for the discrepancy in the measurement of the ratio of electric to magnetic proton form factor, measured through the Rosenbluth separation technique~\cite{Andivahis:1994rq,Arrington:2003tq,Christy:2004rc} and polarization transfer experiments~\cite{Jones:1999rz,Gayou:2001qd,Punjabi:2005wq}, is due to a two-photon exchange process. Such process gives significant correction to the result of Rosenbluth technique, while it gives a relatively small correction to the polarization transfer method. 

The usual strategy in analyzing experimental $e$-$p$ elastic scattering data is to remove the classic radiative corrections~\cite{oldyennie,oldtsai,Mo:1968cg,Maximon:2000hm}, most often using the Mo and Tsai expressions~\cite{Mo:1968cg}, and present the data in a form to be directly compared to the lowest order expressions.  For the differential cross section, one has the Rosenbluth expression
\ba													\label{eq:reduced}
{d\sigma_{Born} \over d\Omega_{lab}} = 
				{ \tau \sigma_R  \over \varepsilon(1+\tau)}
				{d\sigma_{NS} \over d\Omega_{lab}}  \ ,
\ea
where the ``no structure'' cross section $\sigma_{NS}$ is well-known, $\tau \equiv Q^2 / (4 M^2)$, and  $\varepsilon=(1+2(1+\tau)\tan^2\theta/2)^{-1}$, with $M$ being the proton mass and $\theta$ the electron lab scattering angle.  The leading order reduced cross section is given in terms of the magnetic and electric form factors,
\be
\sigma_R^{Born}  = G_M^2(Q^2) + \frac{\varepsilon}{\tau}  G_E^2(Q^2)   \,.
\ee
The reduced cross section at fixed $Q^2$ is linear in $\varepsilon$, and allows a separation of the two form factors.

Additional, hard two-photon, corrections can change the normalization and $\varepsilon$ slope of the reduced cross section, and also introduce terms that are not linear in $\varepsilon$.  At high $Q^2$, the $G_E^2$ contribution to the reduced cross section is small, so that $\varepsilon$ dependent corrections to the big terms have a significant impact on the extraction of $G_E/G_M$.  Calculations of these effects underlie the recent works~\cite{Guichon:2003qm,Blunden:2003sp,Chen:2004tw,Afanasev:2005mp,Blunden:2005ew,Kondratyuk:2005kk} that at least partly reconcile the Rosenbluth and polarization transfer measurements of $G_E/G_M$.  The same calculations also give non-linearity in $\varepsilon$ to the Rosenbluth plot.  Experimentally~\cite{Tvaskis:2005ex,Tomasi-Gustafsson:2004ms}, the upper bounds on the non-linearity are tightening, and it has been noted that the bounds are now in the vicinity of what is expected from theoretical calculations.  One can, as has been done~\cite{Tvaskis:2005ex}, estimate the nonlinearity induced by the two-photon corrections from published plots.   However, we wish here to precisely quantify the non-linearity directly from the calculation of the hard two-photon corrections, in particular, quantify it in two-photon calculations where the intermediate hadronic state is treated at a quark level~\cite{Chen:2004tw,Afanasev:2005mp}, described using generalized parton distributions.

Recently, Chen {\it et al.}~\cite{Chen:2007ac} have also considered nonlinearity in the Rosenbluth plot.  Chen {\it et al.} parameterize the two-photon contributions to the reduced cross section using functional forms that satisfy requirements based on crossing symmetry~\cite{Rekalo:2003xa}.  The values of their parameters they obtain by fitting directly to data, and they show that the two-photon corrections can significantly change the $\varepsilon$-slope of the reduced cross section while introducing only a modest nonlinearity.  We here are doing a more specific calculation using the GPD based model, and get explicit predictions as to what the nonlinearity should be, in the context of this model.  We can also analyze the sensitivity of the predictions to different model GPDs and to different analytic representations of $G_E$ or $G_M$.  

In the next section, after a few remarks about how two-photon exchange was treated farther in the past, we will record how the formulas for the reduced cross section are modified by the hard two-photon process, and then continue with a description of how to specify quantitatively the non-linearity.  In the third and last section we will show calculated results for the nonlinearity parameter, and make some closing remarks.

\begin{figure}[h]
\begin{center}
\includegraphics[width = 2.6in]{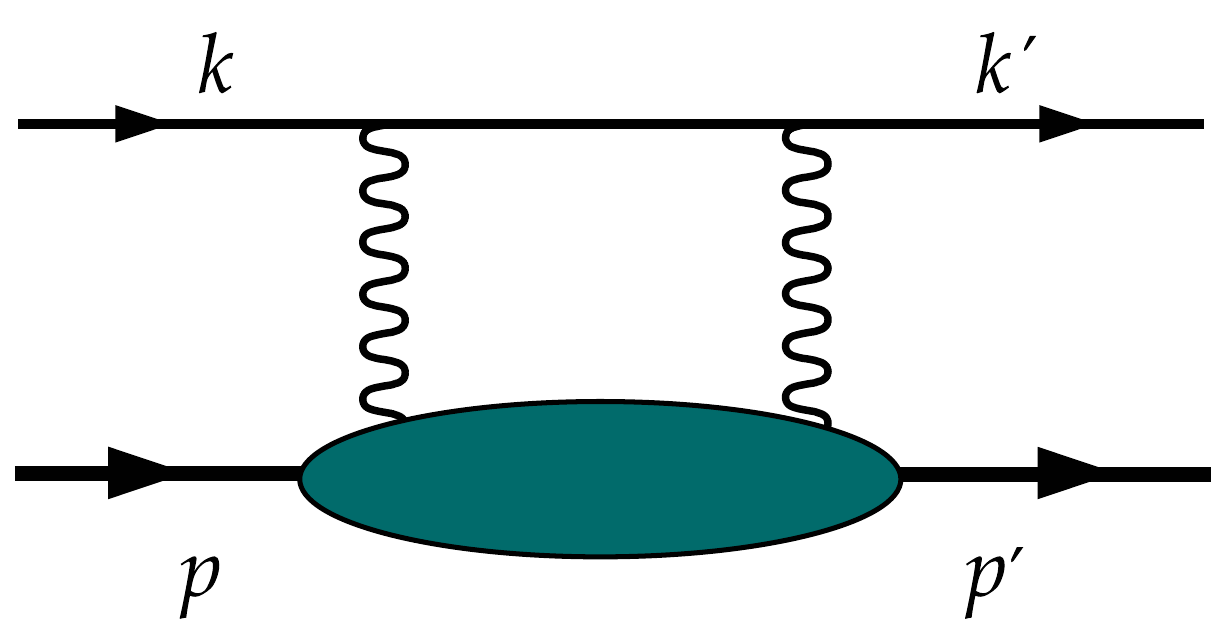}
\caption{Two-photon exchange contribution}
\vskip -4 mm
\label{tpe}
\end{center}
\end{figure}

%
\section{Radiative corrections and formulas}
%

Most radiative corrections to electron-proton elastic scattering were calculated long ago~\cite{oldyennie,oldtsai,Mo:1968cg,Maximon:2000hm}.  Notably not calculated in the far past were the full two-photon exchange contributions, Fig.~\ref{tpe}.  Since energy transferred by one photon can be returned by the other, it is possible for both photons to be quite energetic and hence probe proton structure quite deeply.  The proton structure knowledge needed to complete the two-photon calculation was not available then but is available now, at least to the extent of having sufficient knowledge of proton form factors and/or generalized parton distributions that one can and should attempt a full calculation.  At least two groups have explicitly done so~\cite{Chen:2004tw,Blunden:2005ew}.

The two-photon corrections were not completely neglected in the past.  If one or the other photon is soft, there will be infrared (IR) divergences, and these IR divergences must be calculated and included because they cancel other IR divergences coming from bremsstrahlung corrections.  Calculations of the two-photon corrections in the classic papers were hence done with approximations to the numerators of the expressions that are exact when one photon or the other is soft, but in general amounted to neglecting in the numerator terms of order of the loop momentum-squared.  Thus, although there was an explicitly expressed hope that the neglected contributions would be small~\cite{oldtsai}, there is the real possibility that they could be large.

We turn to a summary of the modern two-photon results following~\cite{Afanasev:2005mp}.  Let $\sigma_R$ be the measured cross section after the classic Mo-Tsai corrections have been applied.  With the additional two-photon corrections,
\ba
\sigma_R
	= \left( G_M^2(Q^2) + \frac{\varepsilon}{\tau} G_E^2(Q^2) \right)
		( 1 + \pi\alpha )
	+ \sigma_{R,hard} \,.
\ea
The hard corrections are
\ba
\sigma_{R,hard} 
	&=& (1 + \varepsilon) \, G_M \, \Re A
			\nonumber	\\ 
	&+& \frac{ \sqrt{2\, \varepsilon \, (1 + \varepsilon)}}{\tau} \, 
			G_E \, \Re B
			\nonumber 	\\
	&+& (1 - \varepsilon) \, G_M \, \Re C\,,
\ea
where $A$, $B$, $C$, and $\sigma_{R,hard}$ depend upon both $Q^2$ and $\varepsilon$, and the characteristic integrals are
\begin{eqnarray}
A &=& \int_{-1}^1 \frac{dx}{x}     
	\frac{\left[(\hat s - \hat u) \tilde{f}_1^{hard} -
		\hat s \hat u \tilde{f}_3 \right]}{(s - u)} 
			\sum_q e_q^2 \, \left( H^q + E^q \right), 
												\nonumber \\
B &=& \int_{-1}^1 \frac{dx}{x}     
	\frac{\left[(\hat s - \hat u) \tilde{f}_1^{hard} - 
		\hat s \hat u \tilde{f}_3 \right]}{(s - u)} 
			\sum_q e_q^2 \, \left( H^q - \tau E^q \right), 
												\nonumber \\
C &=& \int_{-1}^1 \frac{dx}{x} \, \tilde{f}_1^{hard} \, 
	\mathrm{sign}(x) \, \sum_q e_q^2 \, \tilde H^q	\,.
												\label{eq:ABC}
\end{eqnarray}
Each of the three GPDs that enter,  $H^q$, $E^q$, and $\tilde H^q$, are evaluated at zero skewedness, for example,
\be
H^q = H^q(x,\xi, Q^2) \stackrel{\to}{=} H^q(x,0, Q^2)	\,.
\ee

\begin{figure}[t]
\begin{center}
\includegraphics[width = 3.32in]{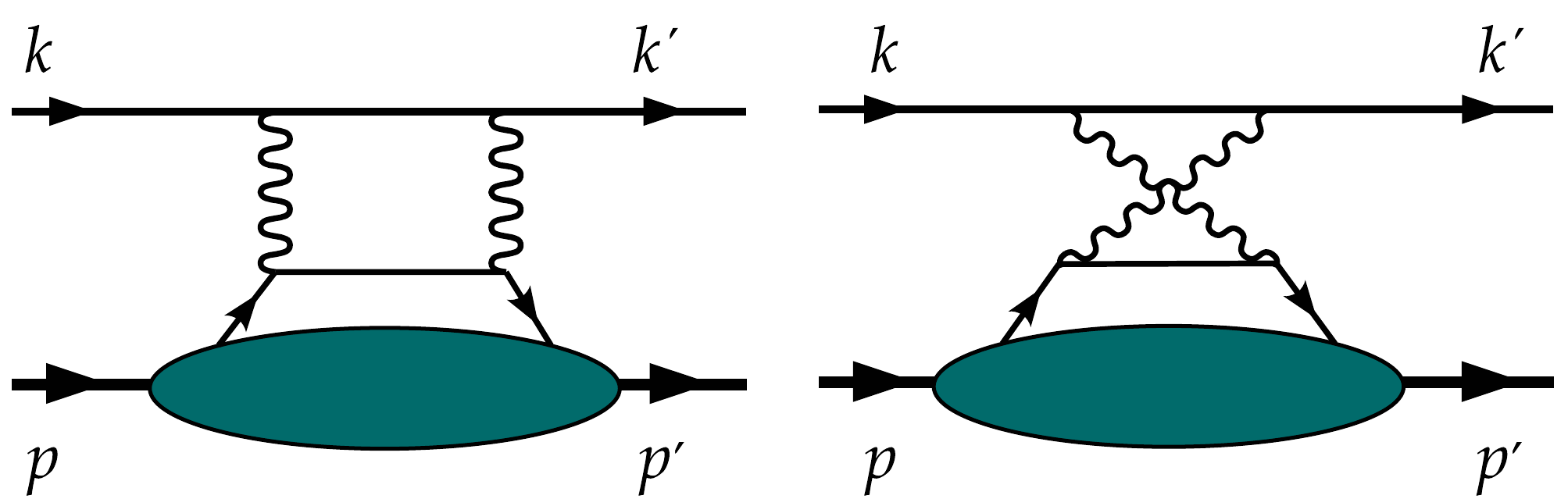}
\caption{Partonic scattering process of direct and crossed box diagram}
\label{tpequark}
\vskip -4 mm
\end{center}
\end{figure}

We picture the two photons interacting with a single quark, as in Fig.~\ref{tpequark}.  The Mandelstam variables for the overall process are $s$, $u$, and $Q^2$, and the evaluation is facilitated by using a frame wherein (external momenta are labeled in Fig.~\ref{tpequark})
\ba
\overline p &\equiv& \frac{1}{2} \left( p'+p \right) = \left( p^+,p^-,p_\perp \right)
									\nonumber \\
	&=& \left( p^+, (M^2 + Q^2/4)/p^+ , 0_\perp \right),
									\nonumber \\
\overline k &\equiv& \frac{1}{2} \left( k'+k \right) = \left( \eta p^+, Q^2/(4\eta p^+), 0_\perp \right),
									\nonumber \\
q &=& p' - p = k - k' = \left( 0,0,q_\perp \right),									
\ea
and one calculates
\be
\eta = \frac{s - u - 2 \, \sqrt{M^4 - s \, u} }{Q^2 + 4 \, M^2}	\,.
\ee
The momentum fraction $x$ in the integrals for $A$, $B$, and $C$ is for the active quark in the proton, and the Mandelstam variabes for the electron-quark subprocess are $\hat s$, $\hat u$, and $Q^2$ with, 
\begin{eqnarray} 
\hat s = \frac{(x + \eta)^2}{4 \, x \, \eta} \, Q^2 \, , \hspace{5mm} 
\hat u = - \frac{(x - \eta)^2}{4 \, x \, \eta} \, Q^2 .
\end{eqnarray}
The elementary electron-quark amplitudes are obtained from electron-muon amplitudes~\cite{VanNieuwenhuizen:1971yn}, and their real parts are~\cite{Afanasev:2005mp}
\ba
\Re \tilde{f}_1^{hard} &=& \frac{\alpha}{ \pi } 
\bigg\{ 
	\frac{1}{2}  \ln  \left( \frac{\hat s}{-\hat u}  \right)  
									\nonumber \\	
&& \hskip -5 mm + \   \frac{Q^2}{4}  \bigg[ \frac{1}{\hat u} \ln^2 \left( \frac{\hat s}{Q^2}  \right)
 - \  \frac{1}{\hat s} \ln^2 \left( \frac{ -\hat u}{Q^2} \right)
	- \frac{1}{\hat s} \pi^2  
		\bigg] 
\bigg\}	,
									\nonumber \\
\Re \tilde{f}_3
&=& \frac{\alpha}{ \pi } \, \frac{1}{\hat s  \hat u} \,
	\bigg\{  \hat s  \ln \left( \frac{\hat s}{Q^2}  \right)  + 
		\hat u \ln \left( \frac{-\hat u}{Q^2}  \right)
									\\ \nonumber
&& \hskip -7 mm + \   \frac{\hat s - \hat u}{2} 
	\left[ \frac{\hat s}{\hat u} \ln^2 \left( \frac{\hat s}{Q^2}  \right)  
		- \frac{\hat u}{\hat s} \ln^2 \left( \frac{-\hat u}{Q^2}  \right)
- \frac{\hat u}{\hat s} \pi^2
\right] 
\bigg\}
\ea

Given a GPD, one can now calculate the two-photon corrections.  However, the GPDs are not at this time definitively known.  We evaluate with two model GPDs, named the ``gaussian'' and the ``modified Regge''~\cite{Guidal:2004nd}, whose parameters are quoted in Ref.~\cite{Afanasev:2005mp}.  Both are constrained to give the quark distribution functions in the appropriate limit, and fit as well as they can the form factors when appropriately integrated.  The modified Regge GPD gives a better fit to the form factors, so would be the current best choice if one chose just one GPD.

\begin{figure}[t]
\begin{center}

\vskip -2.8 cm
\hglue -1.2 cm
\includegraphics[width = 4.3in]{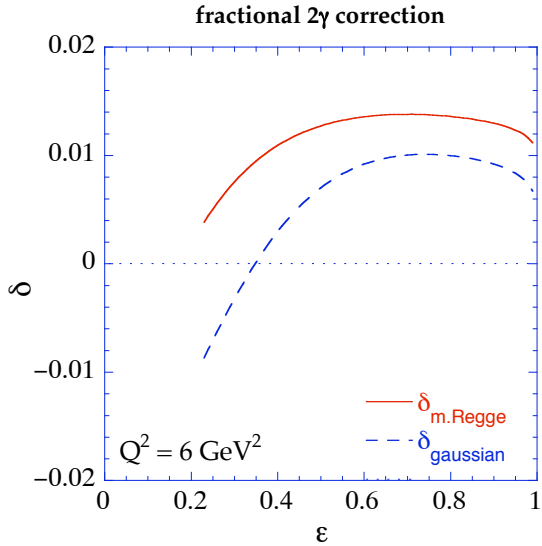}

\vskip -4 cm
\caption{Fractional correction $\delta = (\sigma_R - \sigma_R^{Born} )/ \sigma_R^{Born}$ for two model GPDs and $Q^2 = 6$ GeV$^2$.}
\label{curvature}
\end{center}
\end{figure}

The curvature of the reduced cross section is seen in Fig.~\ref{curvature}, which shows
\be
\delta(\varepsilon,Q^2) = \frac{\sigma_R - \sigma_R^{Born} } { \sigma_R^{Born} }
\ee
for $Q^2 = 6$ GeV$^2$.  Since a partonic calculation is valid at high momentum transfer, we require $-u > M^2$, which leads to the low-$\varepsilon$ cutoff on our curves.

To quantify the nonlinearity, we, following~\cite{Tvaskis:2005ex}, fit the reduced cross section to a quadratic polynomial in $\varepsilon$.  With the $Q^2$ dependence tacit,
\be
\label{quadratic}
\sigma_R(\varepsilon) = p_0 \left( 1 + p_1 \left( \varepsilon - \frac{1}{2} \right)
					+ p_2 \left( \varepsilon - \frac{1}{2} \right)^2  \right) \,,
\ee
where the $p_i$ are constant in $\varepsilon$ (though functions of $Q^2$).  If the curve is not quadratic, the $p_i$ will depend on the interval fitted.  We will, at any $Q^2$, fit to the region $\varepsilon_{\rm low}$ to $1$, where $\varepsilon_{\rm low}$ is fixed by the boundary $-u=M^2$, or
\be
\varepsilon_{\rm low} = \frac{ M^2 }{ M^2 + Q^2/2 }	\,.
\ee

Since a theoretical calculation yields a continuous curve, we can fit using a three-point gaussian method, the same fit method that underlies two-point gaussian integration.  By way of reminder, if ones has a function $f(x)$ defined for the interval $-1 \le x \le 1$, one can expand the function up to the second Legendre polynomial,
\be
f(x) = c_0 P_0(x) + c_1 P_1(x) + c_2 P_2(x)	\,.
\ee
Noting that $P_1$ is zero for $x=x_1=0$ and that $P_2$ is zero for $x = x_{\pm 2} = \pm 1/\sqrt{3}$, one can solve for the $c_i$ in terms of $f(x)$ evaluated at three points,
\ba
c_0 &=& \frac{1}{2} \left( f(x_2) + f(x_{-2}) \right)	\,,		\nonumber \\
c_1 &=& \frac{\sqrt{3}}{2}  \left( f(x_2) - f(x_{-2}) \right)	\,,	\nonumber \\
c_2 &=& f(x_2) + f(x_{-2}) - 2f(x_1)	\,.
\ea
As a side note, the gaussian integral is the integral over the Legendre polynomial fit,
\be
\int_{-1}^1 dx\ f(x) = 2 a_0 \,.
\ee

For a fit interval $a \le \varepsilon \le b$, one maps the above procedure linearly using
\ba
\label{map}
\varepsilon = \frac{b+a}{2} + \frac{b-a}{2} \,x \quad {\rm or} \quad 
	x = \frac{1}{b-a} \left( 2\varepsilon -b -a \right),
\ea
The expansion maps to
\ba
\sigma_R(\varepsilon) = a_0 P_0(x) + a_1 P_1(x) + a_2 P_2(x)	\,,
\ea
where $x$ is given by Eq.~(\ref{map}). The coefficients become
\ba
c_0 &=& \frac{1}{2} \left( \sigma_R(\varepsilon_2) + \sigma_R(\varepsilon_{-2}) \right)	\,,
				\nonumber \\
c_1 &=& \frac{\sqrt{3}}{2}  \left( \sigma_R(\varepsilon_2) - \sigma_R(\varepsilon_{-2}) \right)	\,,
				\nonumber \\
c_2 &=& \sigma_R(\varepsilon_2) + \sigma_R(\varepsilon_{-2}) - 
			2\sigma_R(\varepsilon_1)	\,,
\ea
where
\ba
\varepsilon_1 &=& \frac{b+a}{2}		\,,	\nonumber \\
\varepsilon_{\pm 2} &=& \frac{1}{2} \left (b+a - (b-a)/\sqrt{3} \right)	.
\ea

Simple algebra relates the coefficients $p_i$ for the expansion in $(\varepsilon -1/2)$ of Eq.~(\ref{quadratic}) to the $c_i$ just given,
\ba
p_2 &=&  \frac{1}{p_0} \,  \frac{6 c_2}{ (b-a)^2 }		,	\nonumber \\
p_1 &=& \frac{1}{p_0} \left( \frac{2c_1}{b-a} + \frac{6c_2 (1-b-a)}{(b-a)^2} \right) ,	\\ \nonumber
p_0 &=& c_0  + \frac{1-b-a}{b-a} c_1 
	+ \left[ \frac{3}{2} \left( \frac{1-b-a}{b-a} \right)^2 -  \frac{1}{2} \right] c_2	\,.
\ea
We use $a=\varepsilon_{\rm low}$ and $b=1$, and have a fit procedure that is quite easy to code.

%
\section{Curvature parameter results and conclusions}
%

\begin{figure}[t]

\vskip -1.3 cm
\begin{center}
\includegraphics[width = 3.33in]{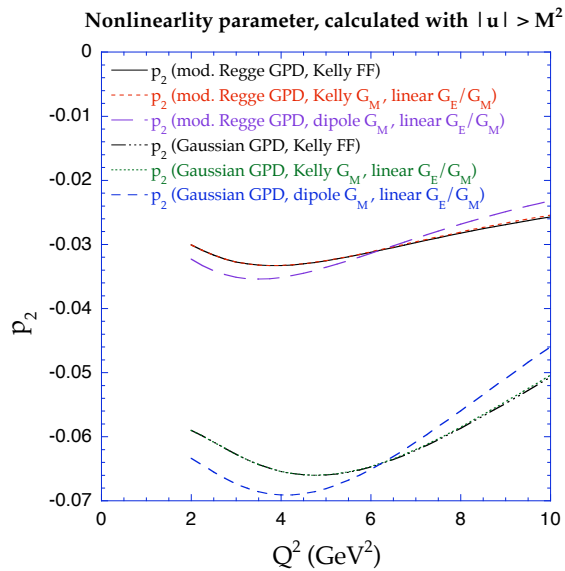}

\vskip -2 cm
\caption{Non-linearity parameter for $\sigma_R \propto 1 + p_1 (\varepsilon-1/2) + p_2 (\varepsilon -1/2)^2$, fitted to calculated $\sigma_R$, with $-u > M^2$.}
\label{curvaturecoeff}
\end{center}
\vskip -4 mm

\end{figure}

The curvature coefficients $p_2$ defined in Eq.~(\ref{quadratic}) that follow from the two GPDs mentioned previously are shown in Fig.~\ref{curvaturecoeff}.

The Figure shows several form factor combinations.  In addition to the Kelly form factors~\cite{Kelly:2004hm}, we there are also plotted two other curves for each GPD.  The second uses the magnetic form factor from Kelly and the electric form factor from the experimenters's linear fit~\cite{Gayou:2001qd},
\be
G_E = \left( 1 - 0.13 \left( Q^2 - 0.04 \right) \frac{G_M}{\mu_p} \right),
\ee
for $Q^2$ in GeV$^2$.  One sees that this change in $G_E$ has little effect.  Using instead the Brash {\it et al.} empirical fit to $G_M$~\cite{Brash:2001qq} gives results barely distinguishable from the previous two curves;  we think the same will be true of any fitted $G_M$ that represents the data well.  For something more extreme, the third curve for each GPD uses the dipole fit for $G_M$ supplemented by the above linear fit for $G_E/G_M$.  Thus the $G_M$ dependence of the Rosenbluth curvature is slight for an up-to-date $G_M$ fit, but the GPD dependence is quite noticeable.

The survey of experimental Rosenbluth data by Tvaskis {\it et al.}~\cite{Tvaskis:2005ex} obtains $\langle p_2 \rangle = 0.019 \pm 0.027$ and a 95\% confidence level upper limit $\left| p_2 \right|_{\rm max} = 0.064$.  These are averages over $p_2$ obtained from many values of $Q^2$, and nearly all of the individual $p_2(Q^2)$ have much larger uncertainty limits.  The most notable individual points are from the $Q^2 = 2.64$ and $3.2$ GeV$^2$ Jefferson Lab measurements which by themselves yield $\langle p_2 \rangle = 0.013 \pm 0.033$, albeit these were measurements where the proton rather than the electron was observed~\cite{Qattan:2004ht}.

Our nonlinearity coefficients calculated using the modified Regge GPDs are negative, and about $1.4$ standard deviations away from the latter points.  One should note again that the GPDs are not yet definitively known, and that the gaussian GPD, yields a significantly larger nonlinearity.    Still, it seems clear that the experiments are at the point of providing a serious test of the two-photon exchange calculations, and in this regard we note that there is a dedicated Rosenbluth nonlinearity test running in Hall C at Jefferson Lab.  We look forward to the analyzed results.

\begin{acknowledgments}
We thank Marc Vanderhaeghen for useful comments and the National Science Foundation for support under grant PHY-0555600.
\end{acknowledgments}

\end{document}